\documentclass[reprint,noshowpacs,noshowkeys,prd,balancelastpage,nofootinbib]{revtex4}
\usepackage{amsfonts}
\usepackage{mdframed}
\usepackage{amssymb}
\usepackage{footnote}
\usepackage{amsmath}
\usepackage{graphicx}
\usepackage{float}
\usepackage[font={footnotesize,it}]{caption}
\usepackage[utf8]{inputenc}
\usepackage{natbib}
\usepackage{tikz}
\usepackage{tikz-3dplot}
\usepackage[colorlinks=true,
            linkcolor=purple,
            urlcolor=purple,
            citecolor=blue]{hyperref}

\begin{document}

\title{Magnetic black hole in Einstein-Dilaton-Square root nonlinear
electrodynamics}
\author{S. Habib Mazharimousavi}
\email{habib.mazhari@emu.edu.tr}
\author{Kanishk Verma}
\email{19500679@emu.edu.tr}
\affiliation{Department of Physics, Faculty of Arts and Sciences, Eastern Mediterranean
University, Famagusta, North Cyprus via Mersin 10, Turkiye}

\begin{abstract}
Starting from the most general action in Einstein-Dilaton-Nonlinear
Electrodynamics (NED) theory, we obtain the field equations. We apply the
field equations for the specific NED known as the square root model coupled
nonminimally to the dilaton scalar field whose self-interaction is in the
Liouville type plus a cosmological constant and solve the field equations.
With a pure magnetic field it is shown that the square root model is the
strong field limit of the Born-Infeld NED theory. In static spherically
symmetric spacetime and a magnetic monopole sitting at the origin, the field
equations are exactly solvable provided the integration constants of the
solutions and the theory constants appearing in the action are linked
through two constraints. As it is known, such an exact solution in the
absence of dilaton i.e., gravity coupled to square root NED doesn't exist.
Therefore, the presence of the dilaton gives additional freedom to solve the
field equations. The obtained spacetime is singular and non-asymptotically
flat and depending on the free parameters it may be a black hole or a
cosmological object. For the black hole spacetime, we study the thermal
stability of the spacetime and show that the black hole is thermally stable
provided its size is larger than a critical value.
\end{abstract}

\date{\today }
\maketitle

\section{Introduction}

The conformally invariant power-law Maxwell nonlinear electrodynamics (NED)
in higher dimensions was introduced by Hassaine and Martinez in \cite{HM1}
where the NED Lagrangian was proposed to be in the form%
\begin{equation}
\mathcal{L}\sim \mathcal{F}^{\frac{d}{4}}  \label{I1}
\end{equation}%
in which $\mathcal{F}=F_{\mu \nu }F^{\mu \nu }$ is the electromagnetic
invariant and $d$ stands for the dimensions of the spacetime. It is obvious
that in $d=4$ the theory coincides with Maxwell's linear theory. In Ref. 
\cite{HM2}, the Lagrangian (\ref{I1}) was generalized into 
\begin{equation}
\mathcal{L}\sim \mathcal{F}^{q}  \label{I2}
\end{equation}%
in which for a generic configuration $\mathcal{F}$ can be negative which
restricts $q$ to be either an integer number or a rational number with odd
denominators., $q\in \tilde{%
%TCIMACRO{\U{211a}}%
%BeginExpansion
\mathbb{Q}%
%EndExpansion
}$ with%
\begin{equation}
\tilde{%
%TCIMACRO{\U{211a}}%
%BeginExpansion
\mathbb{Q}%
%EndExpansion
}=\left\{ \frac{n}{2p+1},\left( n,p\right) \in 
%TCIMACRO{\U{2124} }%
%BeginExpansion
\mathbb{Z}
%EndExpansion
\times 
%TCIMACRO{\U{2124} }%
%BeginExpansion
\mathbb{Z}
%EndExpansion
\right\} .  \label{I3}
\end{equation}%
Ever since this model has been investigated in several papers (see for
instance \cite{P2,P3,P4,P5,P6,P7,P8,P9,P10,P11,P12,P13,P14}). By employing
an adjustable coupling constant $\beta ,$ in \cite{HM3} the above
restriction on $q$ has been removed and consequently, $q$ can get any real
value except $q\neq \frac{1}{2}.$ The exclusion of $s=\frac{1}{2}$ was also
reported in spacetimes other than spherically symmetric. In \cite{H1} a
static cylindrically symmetric magnetic string solution was obtained in the
theory of gravity coupled to the power-law Maxwell NED where $s=\frac{1}{2}$
resulted in no contribution from the magnetic field in the metric function.
The same is also valid for a rotating black string in the same theory with a
combined electric and magnetic field \cite{H2}. In \cite{MH1} the issue of $%
s=\frac{1}{2}$ was addressed where the action was considered to be%
\begin{equation}
I=\frac{1}{2}\int \sqrt{-g}\left( R-2\Lambda -\alpha \mathcal{F}^{s}\right) {%
d^{4}x.}  \label{I4}
\end{equation}%
Furthermore, a generic static spherically symmetric line element and a
combination of radial electric and magnetic fields were considered which are
given by%
\begin{equation}
ds^{2}=-f\left( r\right) dt^{2}+\frac{1}{f\left( r\right) N\left( r\right)
^{2}}dr^{2}+R^{2}\left( r\right) \left( d\theta ^{2}+\sin ^{2}\theta d\phi
^{2}\right)  \label{I5}
\end{equation}%
and%
\begin{equation}
\mathbf{F}=E\left( r\right) dt\wedge dr+P\sin \theta d\theta \wedge d\varphi
\label{I6}
\end{equation}%
respectively where $P>0$ stands for the magnetic charge. In accordance with
the results obtained in \cite{MH1}, 
\begin{equation}
E\left( r\right) =\frac{P\beta }{NR^{2}\sqrt{R^{4}+\beta ^{2}}}  \label{I7}
\end{equation}%
in which $\beta $ is an integration constant representing the electric
charge. This equation reveals that a black hole solution with a pure radial
electric field doesn't exist. However, such a black hole exists in the case
of a pure magnetic field given by $\beta =0$ such that $N\left( r\right) =1,$
$R\left( r\right) =r$ and 
\begin{equation}
f\left( r\right) =1-\frac{2m}{r}-\frac{\Lambda }{3}r^{2}-\frac{\alpha P}{%
\sqrt{2}}.  \label{I8}
\end{equation}%
This black hole solution is the (A)dS-Schwarzschild black hole with a
deficit angle provided by the magnetic monopole charge. Although in the
framework of power-law Maxwell NED, $s=\frac{1}{2}$ was considered in \cite%
{MH1}, a square root Lagrangian with a pure magnetic field was proposed long
ago by Nielsen and Olesen in string theory \cite{NO} and confining potential
by 't Hoof \cite{GT}. Also, Guendelman et al added this square-root term to
Maxwell's Lagrangian to look for confining electric potential in the
framework of Asymptotically de Sitter and anti-de Sitter black holes \cite%
{G1} (see also \cite{G2,G3}). Very recently, a $2+1+p-$dimensional brane
solution was introduced in gravity coupled to the square-root NED powered by
a uniform magnetic field in $z-$direction \cite{M1}. Technically the
electric and magnetic square-root NED are expressed as $\mathcal{L}%
_{e}=-\alpha \sqrt{-\mathcal{F}}$ and $\mathcal{L}_{m}=-\alpha \sqrt{%
\mathcal{F}}$ in which for energy conditions to be satisfied $\alpha $ is a
positive real coupling constant. It is remarkable to observe that the
well-known Born-Infeld NED 
\begin{equation}
\mathcal{L}=b^{2}\left( 1-\sqrt{1+\frac{\mathcal{F}}{b^{2}}}\right)
\label{I9}
\end{equation}%
in the strong field limit implies%
\begin{equation}
\mathcal{L}\rightarrow b\sqrt{\mathcal{F}}\text{ as }\mathcal{F}\rightarrow 
\text{large}  \label{I10}
\end{equation}%
which is the square-root model.

On the other hand, dilatonic black holes have been the subject of several
research papers since the seminal work of Gibbons and Maeda in \cite{GM1}.
The next paper in this line that has popularized the so-called
Einstein-Maxwell-Dilaton theory was published by Garfinkle, Horowitz, and
Strominger \cite{GHS}. Starting from the four-dimensional low-energy
Lagrangian of the string theory a charged dilatonic black hole was
introduced. We note that while the Schwarzschild black hole with a mass
larger than the Planck mass is a good approximate solution in string theory
except in the vicinity of the singularity, the Reissner-Nordstr\"{o}m black
hole is not even an approximate solution in string theory. Hence, the
presence of the dilatonic field as introduced in \cite{GHS} the
corresponding black hole is a good approximation in string theory. After 
\cite{GHS} and knowing the origin of the dilatonic black hole (see also \cite%
{S1,S2,S3}) and its applications in string theory there have been several
exact black hole solutions and their physical properties in the literature
(see for instance \cite{D1,D2,D3,D4,D5,D6}). In \cite{DS1} Dilaton black
holes in Einstein-power-Maxwell-Dilaton have been studied by Dehghani and
Setare. Interestingly, in \cite{DS1} also the authors excluded the
square-root model. Therefore, in this current study, we would like to
address the square-root NED anew in the framework of gravity coupled to a
Dilaton field. Coupling the square-root NED and a scalar field
simultaneously to gravity gives additional freedom to obtain new black holes
with features different than those in the Einstein-power-Maxwell-Dilaton
theory \cite{DSD1}.

\section{The action and the field equations}

We start with the most general action in the Einstein-nonlinear
electrodynamic Dilaton theory, given by ($8\pi G=1$)%
\begin{equation}
S=\frac{1}{2}\int \sqrt{-g}\left( R-2\partial _{\alpha }\psi \partial
^{\alpha }\psi +\mathcal{L}(X,Y,\psi )-V(\psi )\right) {d^{4}x}  \label{1}
\end{equation}%
in which 
\begin{equation}
X={\frac{1}{4}}F_{\mu \nu }F^{\mu \nu }  \label{2}
\end{equation}%
and%
\begin{equation}
Y={\frac{1}{4}}F_{\mu \nu }{^{\star }}F^{\mu \nu }  \label{3}
\end{equation}%
are the electromagnetic invariants, $\psi \left( x^{\mu }\right) $ is the
Dilaton field, $\mathcal{L}(X,Y,\psi )$ is a generic function of $X,Y,$ and $%
\psi $ and $V(\psi )$ is a self-interacting potential for the Dilaton.
Herein, $F_{\mu \nu }=\partial _{\mu }A_{\nu }-\partial _{\nu }A_{\mu }$ and 
${^{\star }}F^{\mu \nu }=-\frac{1}{2\sqrt{-g}}\epsilon ^{\alpha \beta \mu
\nu }F_{\alpha \beta }$ are the electromagnetic tensor and its dual tensor,
respectively with $A_{\mu }\left( x^{\nu }\right) $ the electromagnetic
gauge potential. Variation of the action with respect to the metric tensor $%
g^{\mu \nu }$ yields Einstein's field equation 
\begin{equation}
G_{\mu }^{\nu }=T_{\mu }^{\nu }  \label{4}
\end{equation}%
in which $G_{\mu }^{\nu }=R_{\mu }^{\nu }-\frac{1}{2}R\delta _{\mu }^{\nu }$
is Einstein's tensor and $T_{\mu }^{\nu }$ is the energy momentum tensor
given by%
\begin{equation}
T_{\mu }^{\nu }=-\partial _{\alpha }\psi \partial ^{\alpha }\psi \delta
_{\mu }^{\nu }+2\partial _{\mu }\psi \partial ^{\nu }\psi -\frac{1}{2}%
V\left( \psi \right) \delta _{\mu }^{\nu }+\frac{1}{2}\left( \delta _{\mu
}^{\nu }\left( \mathcal{L}-Y\mathcal{L}_{Y}\right) -\mathcal{L}_{X}F_{\mu
\lambda }F^{\nu \lambda }\right) .  \label{5}
\end{equation}%
Herein, $\mathcal{L}_{X}=\frac{\partial \mathcal{L}}{\partial X}$ and $%
\mathcal{L}_{Y}=\frac{\partial \mathcal{L}}{\partial Y}.$ Upon a
contraction, (\ref{4}) implies%
\begin{equation}
R=2\partial _{\alpha }\psi \partial ^{\alpha }\psi +2V\left( \psi \right)
-2\left( \mathcal{L}-Y\mathcal{L}_{Y}-\mathcal{L}_{X}X\right)  \label{6}
\end{equation}%
which upon plugging into (\ref{4}), Einstein's field equations can be
coasted in the following more convenient form%
\begin{equation}
R_{\mu }^{\nu }=2\partial _{\mu }\psi \partial ^{\nu }\psi +\frac{1}{2}%
V\left( \psi \right) \delta _{\mu }^{\nu }+\frac{1}{2}\left( \delta _{\mu
}^{\nu }\left( Y\mathcal{L}_{Y}+2X\mathcal{L}_{X}-\mathcal{L}\right) -%
\mathcal{L}_{X}F_{\mu \lambda }F^{\nu \lambda }\right) .  \label{7}
\end{equation}%
Furthermore, variation of the action with respect to the gauge field $A_{\mu
}$ gives the nonlinear Maxwell equation described by%
\begin{equation}
\nabla _{\rho }\left( \mathcal{L}_{X}F^{\mu \rho }+\mathcal{L}_{Y}{^{\star }}%
F^{\mu \rho }\right) =0,  \label{8}
\end{equation}%
which together with the Bianchi identity 
\begin{equation}
\nabla _{\mu }({^{\star }}F^{\mu \nu })=0,  \label{9}
\end{equation}%
govern the rule of electrodynamics in the theory. Finally, we variate the
action with respect to the Dilaton scalar field to get the Dilaton equation
expressed by%
\begin{equation}
\nabla _{\mu }\nabla ^{\mu }\psi =-\frac{1}{4}\partial _{\psi }\left( 
\mathcal{L}(X,Y,\psi )-V\left( \psi \right) \right)  \label{10}
\end{equation}%
where $\partial _{\psi }$ stands for $\frac{\partial }{\partial \psi }.$ The
field equations (\ref{7})-(\ref{10}) are the main equations to be considered
in the following section where we introduce a specific $\mathcal{L}(X,Y,\psi
)$ and $A_{\mu }$ as well as a static generic spherically symmetric
spacetime.

Before going further we would like to add that the field equations (\ref{7}%
)-(\ref{10}) in the absence of the scalar field $\psi $\ and in the limit of
Maxwell's linear theory i.e., $L(X,Y,\psi )=-X,$\ simply reduce to%
\begin{equation}
G_{\mu }^{\nu }=\frac{1}{2}\left( F_{\mu \lambda }F^{\nu \lambda }-\frac{1}{4%
}\delta _{\mu }^{\nu }F_{\alpha \beta }F^{\alpha \beta }\right) ,  \label{R1}
\end{equation}%
and%
\begin{equation}
\nabla _{\rho }F^{\mu \rho }=0,  \label{R2}
\end{equation}%
together with the same Bianchi identity given in (\ref{9}). These are the
standard field equations of Einstein-Maxwell's theory. On the other hand, in
the flat spacetime in the absence of the scalar field, the only remained
field equations are Maxwell's equation (\ref{R2}) as well as the Bianchi
identity (\ref{9}). In other words, the theory falls into the classical
Maxwell theory.

Intentionally, we started with the most general action in the context of
Gravity coupled with nonlinear electrodynamics and scalar field to obtain
the field equations in their generic form. This is due to the variety of the
proposed theories in the same context. Therefore, our field equations in the
appropriate choice of $L(X,Y,\psi )$\ and $V\left( \psi \right) $\ reduce to
the known solutions in the literature. In addition to that, there are new
configurations that are of interest for particular reasons which have not
been studied. In this current study, we present one such coupling between
gravity and nonlinear electrodynamics, known as the square root model, and
the Dilaton scalar field. As we have already mentioned in Introduction, the
strong field limit of Born-Infeld Lagrangian in nonlinear electrodynamics is
the square root model. This model has been considered in several research
works, particularly in the context of confinement potentials \cite{NO,GT}.
Moreover, there has been great interest in the Einstein-Born-Infeld-Dilaton
theory in recent years \cite{R5}. In \cite{D2}, Yazadjiev introduced a
non-asymptotically flat black hole solution in this theory. The action of 
\cite{D2} is given by (\ref{1}) in which $V\left( \psi \right) =0$\ and 
\begin{equation}
\mathcal{L}(X,Y,\psi )=4be^{2\gamma \psi }\left[ 1-\sqrt{1+\frac{e^{-4\gamma
\psi }}{2b}X-\frac{e^{-8\gamma \psi }}{16b^{2}}Y^{2}}\right] .  \label{L1}
\end{equation}%
In the pure electric or magnetic setting $Y^{2}=0$\ upon which in the strong
field regime $L(X,Y,\psi )$\ becomes%
\begin{equation}
\mathcal{L}(X,Y,\psi )\sim \sqrt{X}  \label{L2}
\end{equation}%
which implies that in the strong field regime, the theory and its solutions
reduce to the strong regime of the Einstein-Born-Infeld theory and therefore
the effect of the Dilaton field is absent. Keeping this in mind, here in
this study we consider the strong field regime from the beginning (see the
form of $L(X,Y,\psi )$\ and $V\left( \psi \right) $\ in Eqs. (\ref{22}) and (%
\ref{23}), respectively) and the coupling of the Dilaton field is set to be
nontrivial. Therefore in this aspect, our analysis can be considered
complementary to the traditional Einstein-Born-Infeld-Dilaton theory in the
strong field regime.

\section{The solution}

We start with our spacetime ansatz which is static and spherically symmetric
with the line element%
\begin{equation}
ds^{2}=-f\left( r\right) dt^{2}+\frac{1}{f\left( r\right) }%
dr^{2}+R^{2}\left( r\right) \left( d\theta ^{2}+\sin ^{2}\theta d\phi
^{2}\right)   \label{11}
\end{equation}%
in which $f\left( r\right) $ and $R\left( r\right) $ are the metric
functions to be determined. The nonzero components of the Ricci tensor are
given by%
\begin{equation}
R_{t}^{t}=-\frac{f^{\prime \prime }R+2f^{\prime }R^{\prime }}{2R},
\label{12}
\end{equation}%
\begin{equation}
R_{r}^{r}=-\frac{4R^{\prime \prime }f+f^{\prime \prime }R+2f^{\prime
}R^{\prime }}{2R}  \label{13}
\end{equation}%
and%
\begin{equation}
R_{\theta }^{\theta }=R_{\phi }^{\phi }=-\frac{RR^{\prime }f^{\prime
}+fRR^{\prime \prime }-1+fR^{\prime 2}}{R^{2}}.  \label{14}
\end{equation}%
It is helpful to notice that the following relations can be extracted from
the components of the Ricci tensor%
\begin{equation}
R_{t}^{t}=-\frac{1}{2R^{2}}(f^{\prime }R^{2})^{\prime }  \label{15}
\end{equation}%
and%
\begin{equation}
R_{t}^{t}-R_{r}^{r}=2\frac{R^{\prime \prime }}{R}f.  \label{16}
\end{equation}%
Next, we introduce our electromagnetic field which is a radial pure magnetic
field produced by a magnetic monopole sitting at the origin described by the
following two form field%
\begin{equation}
\mathbf{F}=B(r)R^{2}\sin d\theta \wedge d\phi   \label{18}
\end{equation}%
such that its dual-field reads%
\begin{equation}
{^{\star }}\mathbf{F}=B(r)dt\wedge dr.  \label{19}
\end{equation}%
The Bianchi identity then implies $d\mathbf{F}=0$ which simply yields 
\begin{equation}
B(r)=\frac{P}{R^{2}},  \label{20}
\end{equation}%
where $P$ is an integration constant representing the magnetic charge.
Having a pure magnetic field results in $Y=0$ however 
\begin{equation}
X=\frac{1}{2}B^{2}=\frac{1}{2}\frac{P^{2}}{R^{4}}.  \label{21}
\end{equation}%
Moreover, we introduce our BI-nonlinear electrodynamic Lagrangian model
coupled non-minimally to the Dilaton field given by%
\begin{equation}
\mathcal{L}(X,Y,\psi )=4b\left[ 1-\sqrt{1+\frac{e^{4\lambda \psi }}{2b}X-%
\frac{e^{8\lambda \psi }}{16b^{2}}Y^{2}}\right]   \label{S1}
\end{equation}%
in which $\lambda $ is a real constant and $\psi =\psi \left( r\right) .$
With a pure magnetic field configuration $Y=0$ such that we obtain%
\begin{equation}
\mathcal{L}(X,Y,\psi )\simeq -Xe^{4\lambda \psi }\left( 1-\frac{1}{8}\frac{%
Xe^{4\lambda \psi }}{b}+\mathcal{O}\left( \left( \frac{Xe^{4\lambda \psi }}{b%
}\right) ^{3}\right) \right) \text{ as }\frac{X}{b}\rightarrow 0  \label{T1}
\end{equation}%
and%
\begin{equation}
\mathcal{L}(X,Y,\psi )\simeq b\left( -2\sqrt{\frac{2Xe^{4\lambda \psi }}{b}}%
+4-4\sqrt{\frac{b}{2Xe^{4\lambda \psi }}}+\mathcal{O}\left( \left( \frac{b}{%
2Xe^{4\lambda \psi }}\right) ^{3/2}\right) \right) \text{as }\frac{X}{b}%
\rightarrow \infty .  \label{T2}
\end{equation}%
From (\ref{T1}) and (\ref{T2}) we observe that the weak field limit
corresponding to $\frac{X}{b}\ll 1$ and the strong field limit corresponding
to $\frac{X}{b}\gg 1$ of the Lagrangian (\ref{S1}) are given by%
\begin{equation}
\mathcal{L}(X,Y,\psi )\simeq -e^{4\lambda \psi }X\text{ as }b\gg X,\text{ \ }
\label{S2}
\end{equation}%
and%
\begin{equation}
\mathcal{L}(X,Y,\psi )\simeq -\frac{4}{\sqrt{2b}}e^{2\lambda \psi }\sqrt{X}%
\text{ as }b\ll X,  \label{S3}
\end{equation}%
respectively. Note that $b$ is a positive constant and for the pure magnetic
field ansatz (\ref{18}) $X>0$. Since we are aiming to solve the field
equations in the strong field limit of this theory, our main concern in the
sequel will be the Lagrangian (\ref{S3}). Moreover, with a redefinition of
the Dilaton field i.e., $\psi \rightarrow \psi +\frac{\ln \left( 2b\right) }{%
4\lambda },$ one can absorb the coefficient $\frac{1}{\sqrt{2b}}$ and write%
\begin{equation}
\mathcal{L}(X,Y,\psi )=-4e^{2\lambda \psi }\sqrt{X}  \label{22}
\end{equation}%
where for simplicity we turned $\simeq $ into equality. We would like to add
that the strong field regime refers to $1\ll \frac{X}{b}$ and is not indeed
equivalent to a strong electric or magnetic field. This is important to know
that in the very strong magnetic field produced by neutron stars the
phenomenon of pair production is likely to take place and therefore has to
be taken into account. For pair productions in the strong fields namely
electric or magnetic, we refer to \cite{PP1,PP2,PP3} and the references
therein, however, in this study, we assume the fields are weaker than the
threshold fields such that no pair production takes place.

Having, $\psi =\psi \left( r\right) $ and $X=X\left( r\right) $ guarantees
that the nonlinear Maxwell equation is trivially satisfied. On the other
hand, we introduce a Liouville-type self-interacting potential with a
cosmological constant, given by%
\begin{equation}
V\left( \psi \right) =V_{0}e^{2\alpha \psi }+\Lambda  \label{23}
\end{equation}%
such that the Dilaton field equation becomes%
\begin{equation}
\left( R^{2}f\psi ^{\prime }\right) ^{\prime }=R^{2}\left( 2\lambda
e^{2\lambda \psi }\sqrt{X}+\frac{1}{2}\alpha V_{0}e^{2\alpha \psi }\right) .
\label{24}
\end{equation}%
Herein, $V$ and $\Lambda $ are two constants. In addition to (\ref{24}) we
find also Einstein's equations explicitly given by

\begin{equation}
R_{t}^{t}=\frac{1}{2}V,  \label{25}
\end{equation}

\begin{equation}
R_{r}^{r}=2f\psi ^{\prime }{}^{2}+\frac{1}{2}V,  \label{26}
\end{equation}%
and

\begin{equation}
R_{\theta }^{\theta }=\frac{1}{2}\left( V+\frac{4Pe^{2\lambda \psi }}{\sqrt{2%
}R^{2}}\right) .  \label{27}
\end{equation}%
Using (\ref{15})-(\ref{16}) the first two components of Einstein's field
equations are simplified significantly as follows%
\begin{equation}
(f^{\prime }R^{2})^{\prime }=-VR^{2}  \label{28}
\end{equation}%
and%
\begin{equation}
\frac{R^{\prime \prime }}{R}=-\psi ^{\prime 2}.  \label{29}
\end{equation}%
Hence, we are left with 3+1 equations i.e., Eqs. (\ref{27})-(\ref{29}) and (%
\ref{24}) however after some manipulation one can prove that Eq. (\ref{24})
is not independent and results in only three equations (\ref{27})-(\ref{29})
and three unknown functions i.e., $f\left( r\right) ,$ $R\left( r\right) $
and $\psi \left( r\right) .$ Following the traditional ansatz, we suppose 
\begin{equation}
R\left( r\right) =R_{0}e^{\lambda \psi }  \label{31}
\end{equation}%
upon which (\ref{29}) yields%
\begin{equation}
\psi \left( r\right) =\psi _{0}+\frac{\lambda }{1+\lambda ^{2}}\ln r
\label{32}
\end{equation}%
in which $\psi _{0}$ is an integration constant. Considering (\ref{32}) into
(\ref{31}), we also find%
\begin{equation}
R\left( r\right) =R_{0}e^{\lambda \psi _{0}}r^{\frac{\lambda ^{2}}{1+\lambda
^{2}}}.  \label{33}
\end{equation}%
Finally, we consider Eq. (\ref{27}) which is a first-order differential
equation for $f\left( r\right) $ and obtain%
\begin{equation}
f\left( r\right) =Cr^{\frac{1-\lambda ^{2}}{1+\lambda ^{2}}}-\frac{V_{0}}{2}%
\left( 1+\lambda ^{2}\right) e^{-2\lambda \psi _{0}}r^{\frac{2}{1+\lambda
^{2}}}-\frac{\left( 1+\lambda ^{2}\right) \Lambda }{2\left( 1+3\lambda
^{2}\right) }r^{2}  \label{34}
\end{equation}%
where in order to satisfy (\ref{25}) we had to set $\alpha =-\lambda $ and
the following constraint has to be held%
\begin{equation}
P=\frac{\Lambda \sqrt{2}}{2V_{0}\left( \lambda ^{4}-1\right) }  \label{35}
\end{equation}%
and%
\begin{equation}
R_{0}^{2}=\frac{2}{V_{0}\left( 1-\lambda ^{2}\right) },\text{ \ \ \ }\lambda
\neq \pm 1.  \label{36}
\end{equation}%
As it is observed from (\ref{33}) and (34), $V_{0}e^{-2\lambda \psi _{0}}$
may be set as a new $V_{0}$ without any consequence which suggests that $%
\psi _{0}$ can be set to zero. Therefore the metric function (\ref{34})
contains three theory parameters namely $V_{0},\lambda $, and $\Lambda $,
and one integration constant $C$. There are two other integration constants
i.e., $P$ and $R_{0}$ which are not independent of the theory parameters
through Eqs. (\ref{35}) and (\ref{36}). In the limit $\lambda \rightarrow 0$
the action and the line element become%
\begin{equation}
S=\frac{1}{2}\int \sqrt{-g}\left( R-4\sqrt{X}-\left( V_{0}+\Lambda \right)
\right) {d^{4}x}  \label{37}
\end{equation}%
and%
\begin{equation}
ds^{2}=-\left( Cr-\left( \frac{1}{R_{0}^{2}}+\frac{\Lambda }{2}\right)
r^{2}\right) dt^{2}+\frac{1}{Cr-\left( \frac{1}{R_{0}^{2}}+\frac{\Lambda }{2}%
\right) r^{2}}dr^{2}+R_{0}^{2}\left( d\theta ^{2}+\sin ^{2}\theta d\phi
^{2}\right)  \label{38}
\end{equation}%
respectively in which%
\begin{equation}
P=-\frac{\sqrt{2}\Lambda R_{0}^{2}}{4}.  \label{39}
\end{equation}%
Clearly, none of $\Lambda $ and $V_{0}\left( =\frac{2}{R_{0}^{2}}\right) $
can be set to zero i.e., $\Lambda $, $V_{0}\neq 0.$ The spacetime (\ref{38})
possesses a topology similar to the Bertotti-Robinson regular spacetime
given by $AdS_{2}\times S^{2}$ and is regular. To save the generic solution,
with $V_{0}>0$ and $V_{0}<0$ one should impose $\left\vert \lambda
\right\vert <1$ and $1<\lambda $, respectively. As we have already excluded $%
\lambda =\pm 1,$ it admits no solution satisfying all equations. Please note
that the solutions (\ref{33})-(\ref{35}) are very particular because the
solution parameters i.e., the magnetic charge $P$ and $R_{0}^{2}$ are finely
tuned in terms of the Dilaton constant $\lambda $ and the coupling constant $%
V_{0}$. The case $\lambda =\pm 1$ doesn't obey the same pattern and
therefore has to be excluded.

\section{The nature of the solution}

To obtain an overview of the solution, we apply a coordinate transformation
described by 
\begin{equation}
r\rightarrow \left( \frac{R}{R_{0}}\right) ^{\frac{1+\lambda ^{2}}{\lambda
^{2}}}  \label{40}
\end{equation}%
which yields%
\begin{multline}
ds^{2}=-\left( \frac{R}{R_{0}}\right) ^{\frac{2}{\lambda ^{2}}}\left(
1+\lambda ^{2}\right) \left( \tilde{C}\left( \frac{R}{R_{0}}\right) ^{-\frac{%
\lambda ^{2}+1}{\lambda ^{2}}}-\frac{V_{0}}{2}-\frac{\Lambda }{2\left(
1+3\lambda ^{2}\right) }\left( \frac{R}{R_{0}}\right) ^{2}\right) dt^{2}+
\label{41} \\
\frac{\left( 1+\lambda ^{2}\right) dR^{2}}{\lambda ^{4}R_{0}^{2}\left( 
\tilde{C}\left( \frac{R}{R_{0}}\right) ^{-\frac{\lambda ^{2}+1}{\lambda ^{2}}%
}-\frac{V_{0}}{2}-\frac{\Lambda }{2\left( 1+3\lambda ^{2}\right) }\left( 
\frac{R}{R_{0}}\right) ^{2}\right) }+R^{2}d\Omega ^{2}  \notag
\end{multline}%
where $\tilde{C}=\frac{C}{\left( 1+\lambda ^{2}\right) }$.

\subsection{$\left\vert \protect\lambda \right\vert <1$, and $V_{0}>0$}

In the section, we restrict our parameters to $\left\vert \lambda
\right\vert <1$, and $V_{0}>0.$ 
\begin{equation}
\tilde{\Lambda}=\frac{\Lambda }{V_{0}\left( 1+3\lambda ^{2}\right) },
\label{42}
\end{equation}%
\begin{equation}
\beta ^{2}=\frac{2\left( 1+\lambda ^{2}\right) }{\lambda ^{4}R_{0}^{2}V_{0}}%
>0,  \label{43}
\end{equation}%
and%
\begin{equation}
\mu =\frac{2\tilde{C}}{V_{0}}R_{0}^{\frac{\lambda ^{2}+1}{\lambda ^{2}}},
\label{44}
\end{equation}%
which yields%
\begin{equation}
ds^{2}=-R^{\frac{2}{\lambda ^{2}}}\left( \frac{\mu }{R^{\frac{\lambda ^{2}+1%
}{\lambda ^{2}}}}-\tilde{\Lambda}R^{2}-1\right) dT^{2}+\frac{\beta ^{2}dR^{2}%
}{\left( \frac{\mu }{R^{\frac{\lambda ^{2}+1}{\lambda ^{2}}}}-\tilde{\Lambda}%
R^{2}-1\right) }+R^{2}d\Omega ^{2}  \label{45}
\end{equation}%
where%
\begin{equation}
t\rightarrow \frac{2R_{0}^{\frac{2}{\lambda ^{2}}}T}{\left( 1+\lambda
^{2}\right) V_{0}},  \label{46}
\end{equation}%
and $\lambda =0$ is excluded. Considering $\tilde{C}>0,$ with $\tilde{\Lambda%
}>0$ the solution admits a cosmological horizon and a singularity at the
center, however, with $\tilde{\Lambda}<0$ the solution becomes a singular
non-black hole which asymptotically implies%
\begin{equation}
ds^{2}\rightarrow -R^{\frac{2}{\lambda ^{2}}}\left( \left\vert \tilde{\Lambda%
}\right\vert R^{2}-1\right) dT^{2}+\frac{\beta ^{2}dR^{2}}{\left\vert \tilde{%
\Lambda}\right\vert R^{2}-1}+R^{2}d\Omega ^{2}\text{ as }R\rightarrow \infty
,  \label{47}
\end{equation}%
and at the small $R$ limit, it yields%
\begin{equation}
ds^{2}\rightarrow -\mu R^{\frac{1-\lambda ^{2}}{\lambda ^{2}}}dT^{2}+\frac{%
\beta ^{2}}{\mu }R^{\frac{1+\lambda ^{2}}{\lambda ^{2}}}dR^{2}+R^{2}d\Omega
^{2}\text{ as }R\rightarrow \infty .  \label{48}
\end{equation}%
The case where $\tilde{C}<0$ and $\tilde{\Lambda}<0$ results in a singular
black hole with an event horizon with the line element described by 
\begin{equation}
ds^{2}=-R^{\frac{2}{\lambda ^{2}}}\left( \left\vert \tilde{\Lambda}%
\right\vert R^{2}-\frac{\left\vert \mu \right\vert }{R^{\frac{\lambda ^{2}+1%
}{\lambda ^{2}}}}-1\right) dT^{2}+\frac{\beta ^{2}dR^{2}}{\left( \left\vert 
\tilde{\Lambda}\right\vert R^{2}-\frac{\left\vert \mu \right\vert }{R^{\frac{%
\lambda ^{2}+1}{\lambda ^{2}}}}-1\right) }+R^{2}d\Omega ^{2}  \label{49}
\end{equation}%
which asymptotically behaves the same as (\ref{47}). Also, with $\tilde{C}<0$
and $\tilde{\Lambda}>0,$ the solution becomes a dynamic non-black hole in
the sense that the roles of space and time get switched.

\subsection{$\left\vert \protect\lambda \right\vert >1$, and $V_{0}<0$}

In this section, we set $\left\vert \lambda \right\vert >1$ and consequently 
$V_{0}<0.$ Introducing 
\begin{equation}
\tilde{\Lambda}=\frac{\Lambda }{\left\vert V_{0}\right\vert \left(
1+3\lambda ^{2}\right) },  \label{50}
\end{equation}%
\begin{equation}
\beta ^{2}=\frac{2\left( 1+\lambda ^{2}\right) }{\lambda
^{4}R_{0}^{2}\left\vert V_{0}\right\vert }>0,  \label{51}
\end{equation}%
and%
\begin{equation}
\mu =\frac{2\tilde{C}}{\left\vert V_{0}\right\vert }R_{0}^{\frac{\lambda
^{2}+1}{\lambda ^{2}}}>0,  \label{52}
\end{equation}%
which yields%
\begin{equation}
ds_{\pm }^{2}=-R^{\frac{2}{\lambda ^{2}}}\left( 1\pm \frac{\left\vert \mu
\right\vert }{R^{\frac{\lambda ^{2}+1}{\lambda ^{2}}}}-\tilde{\Lambda}%
R^{2}\right) dT^{2}+\frac{\beta ^{2}dR^{2}}{\left( 1\pm \frac{\left\vert \mu
\right\vert }{R^{\frac{\lambda ^{2}+1}{\lambda ^{2}}}}-\tilde{\Lambda}%
R^{2}\right) }+R^{2}d\Omega ^{2}  \label{53}
\end{equation}%
where%
\begin{equation}
t\rightarrow \frac{2R_{0}^{\frac{2}{\lambda ^{2}}}T}{\left( 1+\lambda
^{2}\right) \left\vert V_{0}\right\vert },  \label{54}
\end{equation}%
and $\pm $ stands for positive/negative $\mu .$ With $\mu >0,$ the spacetime
(\ref{52}) may be also a singular interior solution with $\tilde{\Lambda}<0$
or a singular cosmological object with $\tilde{\Lambda}>0.$ Moreover, with $%
\mu <0,$ the spacetime is a black hole with an event horizon with $\tilde{%
\Lambda}<0$ and a black hole with an event and a cosmological horizon with $%
\tilde{\Lambda}<0$ and a proper set of values for the parameters.

\section{Thermal stability of the black hole solution}

\begin{figure}[tbp]
\centering\includegraphics[width=0.5\textwidth]{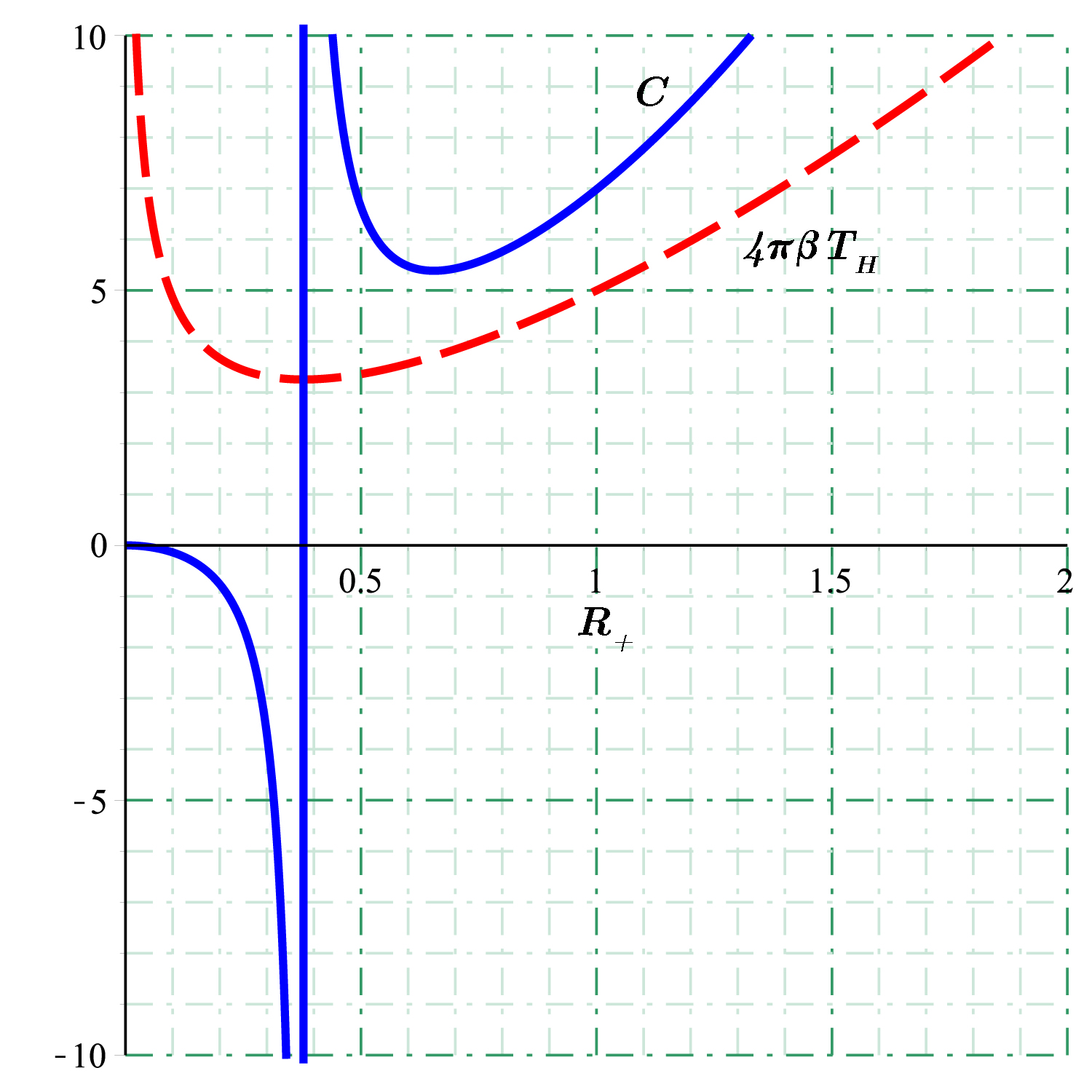}
\caption{Plots of $C$ and $4\protect\pi \protect\beta T_{H}$ in terms of $%
R_{+}$ with $\ell ^{2}=1.$ }
\label{F1}
\end{figure}

In this final section, we investigate the thermal stability of the black
hole solution for $\left\vert \lambda \right\vert >1$ and $V,$ $\mu ,$ $%
\tilde{\Lambda}<0$ such that the line element becomes%
\begin{equation}
ds^{2}=-R^{2\left( \gamma -1\right) }\left( 1-\frac{2M}{R^{\gamma }}+\frac{%
R^{2}}{\ell ^{2}}\right) dT^{2}+\frac{\beta ^{2}dR^{2}}{1-\frac{2M}{%
R^{\gamma }}+\frac{R^{2}}{\ell ^{2}}}+R^{2}d\Omega ^{2},  \label{55}
\end{equation}%
in which $\gamma =\frac{\lambda ^{2}+1}{\lambda ^{2}}$ ($1<\gamma <2$)$,$ $%
\left\vert \mu \right\vert =2M$ and $\left\vert \tilde{\Lambda}\right\vert =%
\frac{1}{\ell ^{2}}.$ Although $M$ is not the mass of the black hole,
writing the spacetime in the form (\ref{55}) gives the impression of having
AdS- Schwarzschild black hole in the limit $\gamma \rightarrow 1$ or
equivalently $\lambda \rightarrow \infty $. First of all this black hole is
non-asymptotically flat and therefore the ADM mass is not defined. Hence we
apply the so-called Brown and York (BY) formalism and obtain the "quasilocal
(QL) mass" of the black hole. In a non-asymptotically flat black hole with
the line-element 
\begin{equation}
ds^{2}=-F\left( R\right) ^{2}dt^{2}+\frac{dR^{2}}{G\left( R\right) ^{2}}%
+R^{2}d\Omega ^{2}  \label{56}
\end{equation}%
the QL mass is defined by%
\begin{equation}
M_{QL}=\lim_{R_{B}\rightarrow \infty }R_{B}F\left( R_{B}\right) \left[
G_{ref}\left( R_{B}\right) -G\left( R_{B}\right) \right]  \label{57}
\end{equation}%
where $G_{ref}\left( R_{B}\right) $ is a non-negative reference function
implying the zero energy of the background spacetime, and $R_{B}$ is the the
radius of the space-like hypersurface. In our case (\ref{55}), one finds $%
F\left( R\right) ^{2}=R^{2\left( \gamma -1\right) }\left( 1-\frac{2M}{%
R^{\gamma }}+\frac{R^{2}}{\ell ^{2}}\right) ,$ $G\left( R\right) ^{2}=\frac{1%
}{\beta ^{2}}\left( 1-\frac{2M}{R^{\gamma }}+\frac{R^{2}}{\ell ^{2}}\right) $
and $G_{ref}\left( R_{B}\right) ^{2}=\frac{1}{\beta ^{2}}\left( 1+\frac{R^{2}%
}{\ell ^{2}}\right) $ which result in 
\begin{equation}
M_{QL}=\frac{M}{\beta }.  \label{58}
\end{equation}%
The line element therefore becomes%
\begin{equation}
ds^{2}=-R^{2\left( \gamma -1\right) }\left( 1-\frac{2\beta M_{QL}}{R^{\gamma
}}+\frac{R^{2}}{\ell ^{2}}\right) dT^{2}+\frac{\beta ^{2}dR^{2}}{1-\frac{%
2\beta M_{QL}}{R^{\gamma }}+\frac{R^{2}}{\ell ^{2}}}+R^{2}d\Omega ^{2}.
\label{59}
\end{equation}%
Next, we calculate the Hawking temperature defined by%
\begin{equation}
T_{H}=\left( -\frac{g_{tt}^{\prime }}{4\pi \sqrt{-g_{tt}g_{rr}}}\right)
_{R=R_{+}}  \label{60}
\end{equation}%
in which $R_{+}$ is the only root of the equation 
\begin{equation}
1-\frac{2\beta M_{QL}}{R^{\gamma }}+\frac{R^{2}}{\ell ^{2}}=0.  \label{61}
\end{equation}%
Hence, we get%
\begin{equation}
T_{H}=\frac{1}{4\pi \beta }R_{+}^{\gamma }\left( \frac{\gamma }{R_{+}^{2}}+%
\frac{\gamma +2}{\ell ^{2}}\right) .  \label{62}
\end{equation}%
Furthermore, by applying the area law of the entropy of black holes i.e., 
\begin{equation}
S=\frac{A}{4}=\pi R_{+}^{2},  \label{63}
\end{equation}%
we calculate the specific heat capacity of the black hole given by%
\begin{equation}
C=T_{H}\frac{\partial S}{\partial T_{H}}=\frac{2\pi R_{+}^{2}\left( \gamma
\ell ^{2}+\left( 2+\gamma \right) R_{+}^{2}\right) }{\gamma \left( \left(
\gamma -2\right) \ell ^{2}+\left( 2+\gamma \right) R_{+}^{2}\right) }.
\label{64}
\end{equation}%
Imposing $1<\gamma <2$, implies that $T_{H}$ in (\ref{62}) is positive
definite however $C$ is positive for $R_{+}>\sqrt{\frac{2-\gamma }{2+\gamma }%
\ell ^{2}}.$ Therefore, the black hole is thermally stable for $R_{+}>\sqrt{%
\frac{2-\gamma }{2+\gamma }\ell ^{2}}$ where both $T_{H}$ and $C$ are
positive. At $R_{+}=\sqrt{\frac{2-\gamma }{2+\gamma }\ell ^{2}}$ there
exists a type-2 phase transition in the sense that the heat capacity
diverges and the temperature becomes minimum (the so-called Davies point).
In Fig. \ref{F1}, we plot $C$ and $4\pi \beta T_{H}$ in terms of $R_{+}$. At
the Davies point the temperature is minimum and the heat capacity diverges.
The stable black hole lies on the right side of the vertical line where both 
$C$ and $T_{H}$ are positive.

\section{CONCLUSION}

We started with the most general action in Einstein-scalar-NED and presented
the field equations. In the static and spherically symmetric spacetime,
these field equations have been solved for the square-root NED coupled to a
Dilaton field accompanied by a Liouville-type self-interacting potential and
a cosmological constant. We recall that in the theory of gravity coupled
with square root NED powered by a magnetic monopole, there is no static
spherically symmetric solution. Here, the presence of the Dilaton field
provided additional freedom such that such an exact solution became
possible. The solution contains overall six constants namely $\lambda
,V_{0},\Lambda ,\psi _{0},P$, and $R_{0}$. While the first three are theory
constants the last three are integration constants. Except for $\psi _{0}$
which was absorbed into $V_{0}$ without loss of generality the other
constants apparently have to satisfy two constraints given in Eqs. (\ref{35}%
) and (\ref{36}). Moreover, with $\lambda =1$ the field equations can not be
solved consistently and consequently has to be excluded. The general
solution is given by the line element (\ref{11}) with the metric functions (%
\ref{33}) and (\ref{34}). The spacetime is singular at the origin where the
magnetic monopole is placed and asymptotically is non-flat. With different
constants configurations, spacetime is either a black hole or a cosmological
object. For a specific setting, the spacetime is a black hole with the line
element given in Eq. (\ref{55}). This solution interestingly coincides with
the Schwarzschild AdS black hole in the limit $\gamma \rightarrow 1$. We
studied the thermal stability of this black hole and showed that it is
stable provided its size is larger than a critical value i.e., $R_{+}>\sqrt{%
\frac{2-\gamma }{2+\gamma }\ell ^{2}}$. Although we haven't considered it in
this study, it is expected that a thermal correction term may be required 
\cite{R1,R2} in the action when the black hole (\ref{59}) is embedded in a
flat Friedmann-Robertson-Walker universe that is known as the McVittie
metric \cite{R3,R4}. We leave further investigation on this matter open for
future work. As a final note, we would like to add that very recently, we
have studied the electric black hole in Einstein-Dilaton-square root
nonlinear electrodynamics \cite{EPJC}. Unlike the magnetic solution
presented in this study, the electric black hole is supported by a zero
self-interacting scalar potential and the electric field is radial and
uniform. The magnetic solution on the contrary is supported by a nonzero
self-interacting potential and the magnetic field is radial but nonuniform.

\textbf{Data Availability Statement:} No data associated with the manuscript.

\textbf{Declarations:} We declare that we have no conflict of interest.

\textbf{Author contributions:} All authors contributed equally to this work.

\end{document}